\documentstyle[12pt, epsf]{article}
\newcommand{\btab}{\begin{tabbing}}
\newcommand{\etab}{\end{tabbing}}

\newcommand{\beqn}{\begin{equation}}
\newcommand{\eeqn}{\end{equation}}
\newcommand{\barr}[1]{\begin{array}{#1}}
\newcommand{\earr}{\end{array}}
\newcommand{\beqna}{\begin{eqnarray}}
\newcommand{\eeqna}{\end{eqnarray}}
\newcommand{\btablec}{\begin{table} \begin{center}}
\newcommand{\etablec}{\end{center} \end{table}}

\newcommand{\gapproxeq}{\lower.7ex\hbox{$\;\stackrel{\textstyle>}{\sim}\;$}}
\newcommand{\plabel}[1]{\label{#1}}
\newcommand{\pbibitem}[1]{\bibitem{#1}}
\def\question#1{{{}}}
\input epsf

\begin{document}
\title{
\begin{flushright} 
\small{hep-ph/9906282}
\\ \small{LA-UR-99-2905}
 \end{flushright} 
\vspace{0.6cm}  
\Large\bf Tensor Glueball--Meson Mixing Phenomenology}
\vskip 0.2 in
\author{Leonid Burakovsky\thanks{\small \em E-mail: burakov@t5.lanl.gov} \
and Philip R. Page\thanks{\small \em E-mail: prp@lanl.gov} \\
{\small \em Theoretical Division, MS-B283, Los Alamos National Laboratory}\\ 
{\small \em P.O. Box 1663, Los Alamos, NM 87545, USA}}
\date{June 1999}
\maketitle
\begin{abstract}
{The overpopulated isoscalar tensor states are sifted using Schwinger--type
mass relations. Two solutions are found: one where the glueball is the
$f_J(2220)$, and one where the glueball is more distributed, with
$f_2(1820)$ having the largest component. The $f_2(1565)$ and $f_J(1710)$
cannot be accommodated as glueball--(hybrid) meson mixtures in the
absense of significant coupling to decay channels.
$f_2^{'}(1525)\rightarrow\pi\pi$ is in agreement with experiment. The
$f_J(2220)$ decays neither flavour democratically nor is narrow.}
\end{abstract}
\bigskip


PACS number(s): \hspace{.2cm}11.15.Tk \hspace{.2cm}12.38.Lg 
\hspace{.2cm}12.39.Mk \hspace{.2cm}12.40.Yx \hspace{.2cm}13.25.Jx 
\hspace{.2cm}14.40.Cs 

\section{Introduction}

The amount of isoscalar tensor states claimed to exist experimentally 
\cite{pdg98} has reached a point where na\"{\i}ve interpretation of these 
states becomes perilous. This is particularly distressing in light of the 
fact that the tensor glueball, a degree of freedom beyond conventional 
mesons, is the second lowest glueball predicted by lattice QCD 
\cite{teper,morn,review} and should be 
manifested in this multiplicity of states. There is a need to bring some 
order by sorting out well understood states. What is sorely needed is 
a model--independent theoretical tool. 

The need for model--independence of the theory is especially prevalent in 
light of the fact that the recently claimed $a_2(1660)$ 
\cite{a21660,ves,argus,l3} has a mass which confounds 
traditional, and often reliable, potential models of radially 
excited P--wave mesons. For example, the difference between the first 
radially excited and ground state isovector $J^{PC}=2^{++}$ (tensor) 
states were predicted to 
be 510 MeV in a relativized quark model \cite{isgur85}, 
while the experimental value is $\sim 340$ MeV. 

In this paper we present a mass--matrix analysis of considerable, although 
not total, generality. Schwinger--type mass formulae \cite{sch} are derived.
It is assumed that the mesons and
glueball mix only via meson--glueball coupling, with no direct meson--meson
coupling.  At any stage of the analysis we restrict ourselves to a finite 
number of mesons (with one glueball) and 
mixing with hypothetical four--quark states is not taken into account.

The term isoscalar ``mesons'' shall refer to the partners of the
light quark isovector mesons, each of which has an $s\bar{s}$ partner.
The isovector mesons will be given labels like P--wave, F--wave or
hybrid meson, indicating the dominant component in a quark model 
interpretation of the state. However, the mass matrix analysis does  not
assume the P--wave, F--wave or
hybrid meson nature of a state, nor that it is an unmixed quark model state.

\section{Masses}

The tensor sector has a few salient features which simplify the analysis.
The first excited lattice QCD tensor glueball is $1.85\pm 0.20$ times
the mass of the scalar glueball \cite{teper}, 
and its effect on the experimental spectrum can hence
safely be neglected. 
This means that we can restrict consideration to the low--lying 
mesons and one primitive (bare) glueball. The mass of the glueball
is reliably estimated by using $M(2^{++})/M(0^{++}) = 1.39\pm 0.04$ 
\cite{morn} or $1.42\pm 0.06$ \cite{teper},
in combination with the average lattice QCD value
$M(0^{++})\approx 1.6$ GeV \cite{scalar}, to be around 2.2 GeV
\cite{review}.
\question{Ref. \cite{morn} estimates
the $2^{++}$ glueball at $2400\pm 25\pm 120$ MeV.}
It is instructive to obtain the lower limit on the tensor glueball mass
allowed by lattice QCD. With 1.5 GeV the lower 
limit for the scalar glueball mass, one obtains a tensor glueball mass 
$\stackrel{>}{\sim }  1.35\;\;\!M(0^{++})\stackrel{>}{\sim } 2.0$ GeV. 
This limit will be
employed later on. 

The isovector tensor mesons should act as beacons for the mass scales of
various nonets. Unfortunately, only the $a_2(1320)$, which we will take
to fix the mass of the primitive $n\bar{n}$ ground state P--wave state, 1P, 
is well established \cite{pdg98}. There is recent 
evidence for $a_2(1660)$ at $1660\pm 40$ MeV or 
$1660\pm 15$ MeV \cite{a21660}, 
for  an $a_2(1600-1700)$ \cite{ves},
evidence at ARGUS for a mainly $2^{++}$ state at 1.7 GeV \cite{argus}
and for an $a_2$ at $1752\pm 21\pm 4$ MeV \cite{l3}. 
The $a_2(1660)$ is taken to fix the mass of 
the primitive $n\bar{n}$ first radially excited P--wave state, 2P.
Additional evidence for the presence of a 2P nonet is provided by its
$1^{++}$ partners. The $a_1 (1700)$ was claimed by BNL \cite{bnl} and 
a similar signal was seen at VES \cite{ves,vesold}.
Recently, weak evidence for $f_1 (\sim 1700)$ was reported \cite{highten}.

\question{Is f1(1700) a hybrid or meson (see BiCEPS and PSS)
a2(1600) in eta pi predicted small by BiCEPS but omega rho large 
consistent with VES.
Evidence for meson nature of a1(1700) in BiCEPS and PSS}

There is also some recent evidence for isovector tensor states at 
 $2060\pm 20$ MeV and $1990^{+15}_{-30}$ MeV \cite{buggf}, signalling 
the 3P and 1F nonets.
Except for these isovector tensor states,
the reasons for expecting the 3P and 1F nonets in a similar mass region
are as follows. There are $f_0(2010)$, $f_0(2060)$  
\cite{pdg98}, and $f_J(2100)$, 
with $J$ most likely 0, at $2115\pm15 \pm 15$ MeV \cite{a21660}.
There are recent indications of an $a_0$ at $2025\pm 30$ MeV  \cite{buggf}.
These $J=0$ states signal P--wave mesons, since neither the ground
state nor the excited scalar glueball is expected in this mass region
\cite{morn}. An $a_1$ at $2100\pm 20$ MeV also indicates 3P.
Given the experimental mass splitting between the 2P and 1P 
nonets noted earlier one also expects the 3P level in this mass region.
We shall take the primitive $n\bar{n}$ 3P level to be at 2.05 GeV.
The 1F nonet is signposted by the $a_4(2050)$,  $f_4(2050)$, $K_4(2045)$,
the recently reported $a_3$ at $1860\pm 20$ MeV \cite{ves} or
$2070\pm 20$ MeV \cite{buggf} and
$f_3$ at $2000\pm 40$ MeV \cite{highten} or $1950\pm 15$ MeV \cite{buggf}. 
There
are recent indications from VES \cite{ves} that the mass of the $a_4(2050)$
is $1944\pm 8\pm 50$ or $1950\pm 20$ MeV. We place the primitive $n\bar{n}$
1F state at 1.94 GeV and the primitive $s\bar{s}$ state higher by 
twice the difference between the $K_4(2045)$ and VES' $a_4(2050)$ masses, 
i.e. at 2.15 GeV. Variation of these masses is discussed in Appendix
\ref{appc}. 

There is some evidence for an isovector tensor state at 
$2265\pm 20$ MeV \cite{buggf}, signalling the 2F or 4P nonets.
The presence of both nonets is indicated by an $f_1$ at $2340\pm 40$ MeV
\cite{highten}, an $a_1$ at $2340\pm 40$ MeV \cite{buggf}, and
an $f_0$ at $2335\pm 25$ MeV \cite{buggf},
which can be 4P but not 2F; or $f_4(2300)$, $K_3(2320)$  \cite{pdg98}, 
an $a_4$ at $2300\pm 20$ MeV \cite{buggf}, an $f_3$ at
$2280\pm 30$ MeV \cite{highten}, and $a_3$ at $2310\pm 40$ MeV, 
which can be 2F but not 4P \footnote{$f_4(2300)$ and $a_4$
may be members of the 1H nonet,
although the nonet appears to be more high--lying, as signalled by the 
$a_6(2450)$ and $f_6(2510)$ \protect\cite{pdg98}.
}.

It is clear that there is no evidence for overpopulation of levels for
isovector tensors, implying that there is no need to introduce a hybrid
meson level\footnote{
The possibility of a tensor hybrid meson in the mass range up to $\sim 2.3$
GeV cannot be excluded theoretically.
Beyond the early MIT bag model estimates, 
constituent gluon models have estimated a tensor hybrid mass, 
most recently at $1.6-1.8$ GeV \protect\cite{kalash}. 
Lattice QCD splittings of hybrid levels indicate that at least for
$b\bar{b}$ hybrids, the tensor hybrid 
is degenerate with the lightest hybrids within errors \cite{manke}. 
However, adiabatic lattice QCD and flux--tube models do not find tensors
on the lowest hybrid adiabatic surface.
Also, tensor mesons are associated with
$0^{++}$ hybrids in bag, constituent gluon and flux--tube models
and adiabatic lattice QCD. There is no indication of an overabundance of
isovector scalar states. 
} up to $\sim 2.3$ GeV. For isodoublet tensors, there is in 
fact an underpopulation: only the well--established $K_2^\ast (1430)$ from 
the 1P nonet, and the marginal $K_2^\ast (1980)$ are known \cite{pdg98}.

To the contrary, there are 13 isoscalar tensor mesons up to $\sim 2.3$ GeV
listed by the Particle Data Group, with 6 
well--established\footnote{Taking both $f_J(1710)$ and $f_J(2220)$ 
to have $2^{++}$ components.} \cite{pdg98}. One expects a glueball, and the 
1P, 2P, 3P and 1F nonets in this mass region, yielding 9 states, and possibly
$n\bar{n}$ 4P and 2F in addition, giving 11 states. There is hence an 
overpopulation of experimental isoscalar tensors, albeit not for the 
well--established ones. 

Since the 1P, 2P, 3P and 1F nonets are expected below 
$\sim 2.3$ GeV, our analysis can safely be 
restricted to a $9\times 9$ mass matrix. There is the possibility of
the 4P and 2F mesons contaminating results at the upper end of our
simulation, at $\sim 2.3$ GeV, which is also investigated.

\section{$5\times 5$ mass matrices}

The mixing of a glueball and $n$ pairs of 
isoscalar mesons is described by the following mass matrix,
motivated in Appendix \ref{appa}, which is 
diagonalized by the masses of $(2n+1)$ physical states: 
\beqn\plabel{mass}
\left(
\begin{array}{cccccccccc}
G & g_1 & g_1\sqrt{2} & g_2 & g_2\sqrt{2} & \cdot  & \cdot  & \cdot  & g_n
 & g_n\sqrt{2}  \\ 
g_1 & S_1 & 0 & 0 & 0 & \cdot  & \cdot  & \cdot  & 0 & 0  \\
g_1\sqrt{2} & 0 & N_1 & 0 & 0 & \cdot  & \cdot  & \cdot  & 0 & 0  \\
g_2 & 0 & 0 & S_2 & 0 & \cdot  & \cdot  & \cdot  & 0 & 0  \\
g_2\sqrt{2} & 0 & 0 & 0 & N_2 & \cdot  & \cdot  & \cdot  & 0 & 0  \\
\cdot  & \cdot  & \cdot  & \cdot  & \cdot  & \cdot  &  &  & \cdot  & \cdot \\
\cdot  & \cdot  & \cdot  & \cdot  & \cdot  &  & \cdot  &  & \cdot  & \cdot \\
\cdot  & \cdot  & \cdot  & \cdot  & \cdot  &  &  & \cdot  & \cdot  & \cdot \\
g_n & 0 & 0 & 0 & 0 & \cdot  & \cdot  & \cdot  & S_n & 0  \\
g_n\sqrt{2} & 0 & 0 & 0 & 0  & \cdot  & \cdot  & \cdot  & 0 & N_n
\end{array}
\right)
\eeqn
\[\hspace{1.2cm}
\Longrightarrow {\rm diag}\;(h_1,\;\!h_2,\;\!h_3\;\!\ldots ,h_{2n},\;\!
h_{2n+1}).
\]
$G$ and $S,N$
stand for the mass of the primitive glueball, and 
$s\bar{s}$ and $n\bar{n}\equiv (u\bar{u}+d\bar{d})/\sqrt{2}$ mesons, 
respectively, the subscript indicating the number of the nonet the state 
belongs to. $h_i$ stand for the masses of the physical states.
$g_i$ are the glueball--meson couplings that have dimensionality 
(mass), in accord with the dimensionality of the diagonal entries
of (\ref{mass}). In what follows, we restrict 
ourselves to the case where the quantities in (\ref{mass}) are real 
numbers\footnote{$g_i$ 
and $-g_i$ gives the same eigenvalues, so we always choose
$g_i$ non--negative.}.
\question{complex matrices? decay?} 

Applying the techniques of ref.  
\cite{sch}, one can obtain $2n$ pairs of relations for the coupling in terms
of the primitive and physical masses:
\[
g_i = \sqrt{-\frac{\prod_{j=1}^{2n+1}(S_i-h_j)}{\prod_{j=1}^{n}(S_i-N_j)
\prod_{j=1\; j\neq i}^{n}(S_i-S_j)}}, \hspace{1cm} i=1,2\ldots n,
\]
\beqn\plabel{sch}
g_i = \sqrt{-\frac{\prod_{j=1}^{2n+1}(N_i-h_j)}{2\prod_{j=1}^{n}(N_i-S_j)
\prod_{j=1\; j\neq i}^{n}(N_i-N_j)}}, \hspace{2.5cm} \mbox{ }
\eeqn
Each pair of these relations represents a Schwinger--type mass formula. Hence, 
for $(2n+1)\times (2n+1)$ mass matrix (\ref{mass}) one has $n$ Schwinger mass 
relations. These $n$ formulae, together with the trace condition for the mass
matrix (\ref{mass}),
\beqn\plabel{trace}
G+S_1+N_1+S_2+N_2+\ldots +S_n+N_n=h_1+h_2+h_3+\ldots +h_{2n}+h_{2n+1},
\eeqn
constitute $n+1$ mass relations for the mixing of a glueball and $n$ meson 
nonets. It is clear that solving such a system of $n+1$ 
mass relations can 
lead to unphysical solutions, e.g., solutions that correspond to all or
some of the couplings being imaginary numbers. Obviously, such solutions will 
not correspond to the initial mass matrix (\ref{mass}), and hence 
should be rejected.

It is difficult to find all the solutions of the Schwinger equations for a 
$9\times 9$ mass matrix of the form (\ref{mass}) numerically. Hence we 
take the approach to solve the Schwinger equations for a $5\times 5$ mass 
sub--matrix, which involves only the primitive and physical masses, 
and then reconstruct the couplings. As all $5\times 5$ sub--matrices
are found, we then obtain the corresponding $9\times 9$ mass matrix.

In this analysis, we pursue the following strategy:

(i) We start with the $5\times 5$ sub--matrix for the glueball and 2P 
and 1F nonets, by fixing the masses of primitive $n\bar{n}$ for 
2P, and $n\bar{n}$ and $s\bar{s}$ for 1F, and obtain the primitive 
2P $s\bar{s}$ mass. We then consider another $5\times 5$ 
sub--matrix for the glueball and 2P and 3P nonets, with fixed: 
both $n\bar{n}$ and $s\bar{s}$ masses of the 2P 
nonet from the previous simulation, and the primitive $n\bar{n}$ mass for the 
3P nonet. 
In both $5\times 5$ sub--matrix simulations we obtain {\it all} the 
solutions of 
the Schwinger equations (\ref{sch}).

(ii) {\it Input:} 
For the two $5\times 5$ sub--matrix simulations, we fix the following
values of the primitive masses (in GeV): $N=1.66$ for 2P, $N=1.94,$ $S=2.15$
for 1F and $N=2.05$ for 3P.
We also take one of the physical states to have a mass in agreement with one
of the glueball candidates (which we review in the next section), 
and the other three 
physical states to have masses in agreement with three states among 
$f_2(1565),$ $f_2(1640),$ $f_J(1710),$ $f_2(1810),$ $f_2(1950),$ $f_2(2010),$ 
$f_2(2150),$ $f_J(2220),$ $f_2(2300)$  and $f_2(2340),$ excluding the state 
already chosen for the physical glueball. 
{\it Output:}
We then solve the system of three 
equations (two Schwinger formulae and the trace condition) for three 
unknowns: 
$G,$ $S$ for 2P and the remaining fifth physical mass for the first 
$5\times 5$ simulation, and $G,$ $S$ for 3P and the remaining fifth
physical mass for the second $5\times 5$ simulation. We require that
the fifth physical mass from each $5\times 5$ simulation is among the
physical states mentioned above.

(iii) Since we take the $f_2(1275)$ and $f_2^{'}(1525)$ as the established 
ground 
state 1P tensor mesons, we incorporate them later in the 
full $9\times 9$ mass matrix analysis.

(iv) We discard the possibility that $f_2(1420)$ exists. Although claimed
by a number of old experiments in a variety of production processes, recent
experiments do not confirm its existence. This is most vividly illustrated
by its observation in (mostly) double Pomeron exchange in 
$pp\rightarrow p_f(\pi^+\pi^-)p_s$ at $\sqrt{s}=63$ GeV
\cite{review,isr}. Recent examination of the {\it same} reaction does not
see any evidence for $f_2(1420)$ \cite{wa10299}.

We admit the following criteria for holding physical solutions and separating 
out non--physical ones:

(i) The output fifth physical mass lies within a mass range allowed 
by data for one 
of the experimental candidates.
 
(ii) The mass of the primitive glueball satisfies $G\geq 2$ GeV.

(iii) In all the cases when a primitive $s\bar{s}$ mass is to be obtained, it
is higher than the corresponding $n\bar{n}$ mass, and the $s\bar{s}-n\bar{n}$
mass splitting is consistent with the quark model motivated estimate 
$200\pm 50$ MeV \cite{scalar}.

\section{$9\times 9$ mass matrices}

Various physical states have been suggested as tensor glueball
candidates in the
literature:

{\bf $f_J(2220)$:}
The $f_J(2220)$ is strongly produced in $J/\psi$ radiative decay,
and not seen in $\gamma\gamma$ collisions, 
suggesting glueball character if $J=2$ \cite{review,farrar}. 
The flavour democratic decay pattern and small total
width of $f_J(2220)$ is also cited
as evidence for its glueball nature \cite{bai96}.

{\bf $f_2(2150)$:}
This was suggested in ref. \cite{prok}.

The nearness of the mass of $f_J(2220)$ and $f_2(2150)$ to the tensor
glueball mass predicted by lattice QCD is often cited as evidence for their
glueball nature \cite{review}.

{\bf $f_2(1950)$:}
The $p_T$ dependence of the $pp$ central production of $f_2(1950)$ is
consistent with its glueball character according the Close--Kirk glueball
filter \cite{review,wa102}. However, it was admitted that the 
structure seen in 
central production may represent more than one resonance \cite{wa102}.

{\bf $f_J(1710)$:}
The glueball nature of this state is suggested by 
its $p_T$ dependence in central production
\cite{wa102} and its production in 
``glue--rich'' $p\bar{p}$ annihilation \cite{farrar}, although its strong 
production in $J/\psi$ radiative decay is consistent with expectations
for $q\bar{q}$ if J=2 \cite{farrar}.

For each $5\times 5$ case, we take one of the above four glueball 
candidates to be one of the five physical states.

Having completed the double $5\times 5$ sub--matrix analysis and 
fixing $N=1.318$ for 1P \cite{pdg98}, the full $9\times 9$ mass matrix is now
recovered by solving the Schwinger equations (\ref{sch}) exactly, using the
solutions obtained for the two $5\times 5$ matrices as initial values for the
search routine. We therefore do {\it not} obtain {\it all} of the
$9\times 9$ mass matrix solutions, but only the ones similar to the
ones found formerly with the two $5\times 5$ matrices. 

There are two different solutions, which
are almost identical with respect to the physical masses.
Particularly, the primitive glueball masses are consistent with
$2.0-2.1$ GeV predicted by recent models
\cite{review,swanson96} and the lattice QCD predictions
mentioned earlier.
The couplings are in the range $30-120$ MeV for the various nonets.
These values are similar to $43\pm 31$ MeV predicted in lattice QCD
for ground state isoscalar scalars \cite{LW1}. 
We find that the physical masses are insensitive to changes in the 
input, but that
the valence content is more sensitive: especially for states at similar masses
to where the parameters are changed, and for small valence components  
(see Appendix \ref{appc}). 

(i) For the first solution, the $f_J(2220)$ turns out to be the physical
glueball. For this solution, the initial $9\times 9$ mass matrix is 
(shown are the values of the primitive masses and couplings rounded to 
the second 
decimal digit; all values are given in GeV)
\beqn\plabel{mass1}
\left(
\begin{array}{ccccccccc}
2.10 & 0.03 & 0.03\;\!\sqrt{2} & 0.04 & 0.04\;\!\sqrt{2} & 0.09 & 
0.09\;\!\sqrt{2} & 0.12 & 0.12\;\!\sqrt{2}  \\
0.03 & 2.28 & 0 & 0 & 0 & 0 & 0 & 0 & 0  \\
0.03\;\!\sqrt{2} & 0 & 2.05 & 0 & 0 & 0 & 0 & 0 & 0  \\
0.04 & 0 & 0 & 2.15 & 0 & 0 & 0 & 0 & 0  \\
0.04\;\!\sqrt{2} & 0 & 0 & 0 & 1.94 & 0 & 0 & 0 & 0  \\
0.09 & 0 & 0 & 0 & 0 & 1.84 & 0 & 0 & 0  \\
0.09\;\!\sqrt{2} & 0 & 0 & 0 & 0 & 0 & 1.66 & 0 & 0  \\
0.12 & 0 & 0 & 0 & 0 & 0 & 0 & 1.55 & 0  \\
0.12\;\!\sqrt{2} & 0 & 0 & 0 & 0 & 0 & 0 & 0 & 1.318
\end{array}
\right) ,
\eeqn
the physical masses are 
\beqn\plabel{phys1}
2.29,\;\;\;2.23,\;\;\;2.14,\;\;\;2.04,\;\;\;1.93,\;\;\;1.82,\;\;\;
1.64,\;\;\;1.52,\;\;\;1.28,
\eeqn
and the valence content of the physical states is
\beqn\plabel{val1}
\left(
\begin{array}{rrrrrrrrr}
\underline{0.38} & {\bf 0.91} &  0.07 & \underline{0.11} & 0.06 & 0.08 
& 0.08 & 0.06 & 0.07  \\
\underline{0.73} & -\underline{0.42} &  \underline{0.18} &  
\underline{0.38} &  \underline{0.14} &  \underline{0.17} &  
\underline{0.17} &  \underline{0.13} &  \underline{0.14}\\
-\underline{0.32} &  0.07 & -\underline{0.16} &  {\bf 0.92} & -0.09 
& -\underline{0.10} & -0.08 & -0.07 & -0.07  \\
-\underline{0.18} & 0.02 &  {\bf 0.97} &  0.07 & -\underline{0.10} 
& -0.08 & -0.06 & -0.05 & -0.04  \\
-\underline{0.15} &  0.01 &  0.05 &  0.03 &  {\bf 0.97} 
& -\underline{0.14} & -0.07 & -0.05 & -0.04  \\
-\underline{0.19} &  0.01 &  0.04 &  0.02 &  \underline{0.10} &  {\bf 0.96} 
& -\underline{0.15} & -0.09 & -0.07  \\
-\underline{0.18} & 0.01 & 0.02 & 0.01 & 0.03 & 0.08 & {\bf 0.94} 
& -\underline{0.25} & -\underline{0.10}  \\
-\underline{0.20} &  0.01 &  0.02 &  0.01 &  0.03 &  0.06 
&  \underline{0.19} &  {\bf 0.94} & -\underline{0.17} \\
-\underline{0.23} & 0.01 & 0.01 & 0.01 & 0.02 & 0.04 & 0.08 
& \underline{0.10} & {\bf 0.96}
\end{array}
\right) .
\eeqn
 
(ii) For the second solution, the the physical
glueball is distributed, with $f_2(1810)$ containing the largest component.  
Although the highest mass state appears to have the largest glueball 
component, we shall see in section \ref{third} that the content changes as 
more high mass states are introduced. For this solution, the initial 
$9\times 9$ mass matrix is
\beqn\plabel{mass2}
\left(
\begin{array}{ccccccccc}
2.05 & 0.10 & 0.10\;\!\sqrt{2} & 0.11 & 0.11\;\!\sqrt{2} & 0.08 & 
0.08\;\!\sqrt{2} & 0.11 & 0.11\;\!\sqrt{2}  \\
0.10 & 2.27 & 0 & 0 & 0 & 0 & 0 & 0 & 0  \\
0.10\;\!\sqrt{2} & 0 & 2.05 & 0 & 0 & 0 & 0 & 0 & 0  \\
0.11 & 0 & 0 & 2.15 & 0 & 0 & 0 & 0 & 0  \\
0.11\;\!\sqrt{2} & 0 & 0 & 0 & 1.94 & 0 & 0 & 0 & 0  \\
0.08 & 0 & 0 & 0 & 0 & 1.95 & 0 & 0 & 0  \\
0.08\;\!\sqrt{2} & 0 & 0 & 0 & 0 & 0 & 1.66 & 0 & 0  \\
0.11 & 0 & 0 & 0 & 0 & 0 & 0 & 1.55 & 0  \\
0.11\;\!\sqrt{2} & 0 & 0 & 0 & 0 & 0 & 0 & 0 & 1.318
\end{array}
\right) ,
\eeqn
the physical masses are 
\beqn\plabel{phys2}
2.38,\;\;\;2.23,\;\;\;2.12,\;\;\;2.01,\;\;\;1.95,\;\;\;1.82,\;\;\;
1.63,\;\;\;1.52,\;\;\;1.28,
\eeqn
and the valence content of the physical states is
\beqn\plabel{val2}
\left(
\begin{array}{rrrrrrrrr}
 \underline{0.64} & \underline{0.58} & \underline{0.27} 
& \underline{0.31} & \underline{0.23} & \underline{0.12} 
& \underline{0.10} & 0.08 & 0.09  \\
 \underline{0.31} & -\underline{0.79} & \underline{0.24} 
& \underline{0.42} & \underline{0.17} & 0.09 & 0.06 & 0.05 & 0.05  \\
-\underline{0.22} & \underline{0.15} & -\underline{0.45} 
& \underline{0.82} & -\underline{0.19} & -\underline{0.11} 
& -0.06 & -0.04 & -0.04 \\
-\underline{0.22} &  0.08 &  \underline{0.75} &  \underline{0.17} 
&-\underline{0.50} &-\underline{0.30} &-0.07 &-0.05 &-0.05  \\
-0.02 &  0.01 &  0.03 &  0.01 &-\underline{0.47} &  \underline{0.88} 
&-0.01 &-0.01 &-0.01  \\
-\underline{0.48} & \underline{0.11} & \underline{0.30} 
& \underline{0.16} & \underline{0.63} & \underline{0.30} 
& -\underline{0.33} & -\underline{0.19} & -\underline{0.15}  \\
-\underline{0.22} &  0.04 &  0.08 &  0.05 &  \underline{0.11} 
&  0.06 &  {\bf 0.91} &-\underline{0.30} &-\underline{0.11} \\
-\underline{0.25} & 0.03 & 0.07 & 0.04 & 0.09 & 0.07 &
 \underline{0.20} & {\bf 0.92} & -\underline{0.19}  \\
-\underline{0.25} & 0.03 & 0.05 & 0.03 & 0.06 & 0.03 & 0.07 
& \underline{0.10} & {\bf 0.96}
\end{array}
\right) .
\eeqn

\section{Discussion of mass matrix results}

Both solutions have the following similarities:

(i) The physical states are $f_2(1270),$ $f_2^{'}(1525),$ $f_2(1640),$ 
$f_2(1810),$ $f_2(1950),$ $f_2(2010),$ $f_2(2150),$ $f_J(2220)$ and either 
the $f_2(2300)/f_2(2340)$ or $f_2(2340)$ as solutions in the 
first and second cases, 
respectively\footnote{Within experimental mass uncertainty 
\protect\cite{pdg98}, we cannot
distinguish between $f_2^{'}(1525)$ or $f_2(1565)$. In section 
\protect\ref{f1525}, we show that
$f_2^{'}(1525)\rightarrow\pi\pi$ is consistent with experiment if it 
is taken to have the
valence content of the second state, which we thus identify as 
$f_2^{'}(1525)$.}. We never found the $f_2(1565)$ and $f_J(1710)$. 
The reason why $f_2(1810)$ is
found instead of these resonances is because the primitive 2P $s\bar{s}$ is 
required to $250\pm 50$ MeV from the input $a_2(1660)$ mass.

(ii) The valence content has  almost entirely the same signs between the
various components, the only exception being different signs for the two 
dominant meson components
in $f_2(1950)$.\question{can one use this to distinguish solutions 
in phenomenology?}

(iii) The physical mesons have a substantial glueball content, 
contrary to na\"{\i}ve expectations, 
with the exception of $f_2(1950)$ in solution 2.
It has been argued phenomenologically that experimental data demand physical
mesons with appreciable glueball content \cite{bz}.
This would, for example, explain why
$f_2(2010),f_2(2300)$ and $f_2(2340)$ were observed in the OZI 
forbidden process
$\pi p\rightarrow \phi\phi n$, and would suggest that several 
tensor mesons should
be produced in glue--rich processes. For example, $f_2(1270)$
was observed in gluon fusion \cite{break}.
The small glueball component in $f_2(1950)$ in solution 2 is
apparently in contradiction with the 
Close--Kirk filter.

(iv) The $f_2(1270),f_2^{'}(1525)$ and $f_2(1640)$ are composed of 
more than 90\% of the expected primitive state.

(v) For the 1P nonet, $S+N=1.55+1.318=2.868$ GeV is
consistent with $2M(K_2^\ast )=2.858\pm 0.01$ GeV \cite{pdg98}.

The solutions differ as follows:

(i) In the first solution all physical mesons have one
component which has a valence content of larger than 90\%, i.e. the state 
is dominantly a specific primitive state.  
For this solution, valence components of physical mesons greater than
10\% only occur within two primitive states of the dominant primitive state. 
The second solution does not behave in this way. 

(ii) For the first solution, the physical glueball has substantial 
valence content in
all the meson states it couples to, contrary to the second solution.

(iii) In the first solution the couplings decrease with increased 
radial excitation
(from 1P to 3P), as one would na\"{\i}vely expect \cite{pertqcd1};
 while for the second solution the
couplings remain approximately the same, as expected from Regge theory 
\cite{ohio}.
An advantage of our method is that all information on 
couplings are predictions, using (\ref{sch}), once the masses are known.

(iv) The first solution has the slight disadvantage
that the primitive $s\bar{s}$--$n\bar{n}$ mass splitting $\sim 180$ MeV 
in the 2P nonet is a little small, versus an agreeable $\sim 290$ MeV for the
second solution.

(v) For the second solution, it 
is seen in (\ref{val2}) that both the 
$f_2(1950)$ and $f_2(2150)$
have large $n\bar{n}$ and $s\bar{s}$ components, and neither 
of them are in destructive interference. Hence, both of these states will have
many different decay modes, which is in excellent agreement with data (these 
different decay modes are amongst 
the main reasons for each of these states to be chosen
as the tensor glueball candidate by different groups). Although this feature 
is absent for these states for the first solution 
(\ref{mass1})-(\ref{val1}), where $f_2(1950)$ is mostly $n\bar{n}$ and
$f_2(2150)$ mostly $s\bar{s}$, many of the observed decay modes arise from 
connected decay of both $u\bar{u}$ and $s\bar{s}$ components, so that this
avenue to distinguish between solutions may not be definitive.

\subsection{$f_2(1565)$ and $f_J(1710)$}

It is possible that $f_2(1565)$ and 
$f_2(1640)$ are aspects of the same state, which would remove one extra 
state.
We have nevertheless 
attempted to find solutions where $f_2(1565)$ and $f_J(1710)$ are 
physical states in addition to the states (\ref{phys1}),(\ref{phys2}),
by adding another nonet to form an $11\times 11$ matrix. Of course, one
can insert primitive states at these masses with an unrealistically small
glueball--meson coupling and an unrealistic $s\bar{s}$--$u\bar{u}$ mass
splitting of 150 MeV and obtain a solution. However, no realistic solutions
are found. 

We have neglected the mixing of mesons 
with decay channels thoughout, since it
is believed to produce only tiny mass shifts \cite{geiger}.
However, scenarios where there is large coupling to 
decay channels, e.g. $\omega\omega,\;\rho\rho,\;
K^{\ast}K^{\ast}$ and $\phi\phi$, have been advanced by various authors.

$f_2(1565)$ decays to $\rho\rho$ and $\omega\omega$ and has an
abnormally small branching ratio to $\pi\pi$ and $\eta\eta$ 
\cite{pdg98}. This, together with the nearness of $f_2(1565)$
to the $\rho\rho$ and $\omega\omega$ thresholds has lead to suggestions
that $f_2(1565)$ is a 
$\rho\rho$ molecule or a baryonium state \cite{torn}. 
Also, $f_2(1565)$ may be the isoscalar
partner of the isotensor tensor enhancement $X(1600)$ \cite{pdg98}, which, if
it is resonant, must be a degree of freedom beyond glueballs and 
(hybrid) mesons.

A more modest suggestion is that
the mass of the 2P $n\bar{n}$ state found in our formalism is shifted
downward by the $\rho\rho$ and $\omega\omega$ thresholds, which 
the $^3P_0$ model predicts it to couple strongly to \cite{biceps}.


$f_J(1710)$ has been suggested as a $K^*K^*$ molecule \cite{dooley}.
However, it is not well established that a $J=2$ component exists.
BES separated both $J=0$ and $J=2$ components, 
with the tensor state having mass $1697$ MeV and a width of $176$ MeV 
\cite{bes1770}. However, recent evidence supports only the $J=0$ component
\cite{wa10299,f1770}. 

\section{$13\times 13$ mass matrix\plabel{third}}

Once the $9\times 9$ mass matrix is fixed, one can easily add extra
meson nonets to it. We add the primitive $n\bar{n}$ and $s\bar{s}$
masses of the 2F nonet at 2.3 and 2.5 GeV, respectively, to the 
mass matrices (\ref{mass1}),(\ref{mass2}). 
Similarly, 
the primitive states of the 4P nonet are added at 2.35 and 2.55 GeV, 
to yield a $13\times 13$ mass matrix. The physical states in Eqs.
(\ref{phys1}),(\ref{phys2}) are required to be among the physical states,
i.e. {\it both} $f_2(2300)$ and $f_2(2340)$.
 
The result is a $13\times 13$ ``counterpart'' to each $9\times 9$ 
matrix  (\ref{val1}),(\ref{val2}), called
solutions 1a and 2a. 
The primitive states in common have the same couplings and primitive masses, 
and similar valence content (with the same signs).
The valence content
of a given primitive state tends to decrease from the
$9\times 9$ counterpart, since the physical state is
spread over more primitive states. 

Remarkably, the ratios of valence contents of the $13\times 13$ solutions
and their $9\times 9$ counterparts remain
extremely similar ($\sim 1\%$), except for the components in the 
$9\times 9$ matrix which has
similar mass to the new components being added.
This means that for low--lying states, there is usually no need 
extend the number
of primitive components in order to study decay.

We also find two new solutions, called 1b and 2b, since they are 
respectively similar to the
$9\times 9$ solutions 1 and 2. They have, however, no 
$9\times 9$ counterparts. Solution 2b is displayed in Appendix \ref{appc}. 

In all $13\times 13$ solutions, three new, experimentally undiscovered, 
physical states appear at masses beyond the $f_2(2340)$.
The dominant
glueball component in solutions 2a and 2b is found in one of the three new
high mass states.

We note that because the number of nonets stable under decay is 
expected to be finite due to pair creation in QCD \cite{bur98}, the largest
mass matrix that need to be analysed is finite. 

\section{Decays}

In order to calculate the decay of a physical state to an exclusive final state
 it is necessary to add the decay amplitudes
of all its primitive components, weighted by their valence content.
This is demonstated in Appendix \ref{appd}.  

The decay amplitudes of the primitive components will be calculated in the
$^3P_0$ model, meaning that  pair creation is with vacuum quantum numbers
and decays proceed via a connected quark diagram. Unless otherwise noted, the
decays are calculated using the ``relativistic'' phase space convention
and parameters of ref. \cite{biceps}. Another convention is
``mock meson'' phase space 
with the parameters of ref. \cite{kokoski87}. In the mass matrix analysis,
specific quark model identifications were not assumed for the various
components. However, to calculate the decays, the quark model content
indicated by the label of a component will be assumed.

The preceding mechanism whereby the 
physical glueball decays via primitive meson components, which is closely
related to the primitive glueball decaying via intermediate primitive mesons
\cite{ohio}, provides the first theoretical understanding of the 
scalar glueball
decay pattern found in lattice QCD \cite{scalar}.

When one combs the experimental data available on isoscalar tensor mesons
more massive than the $f^{'}_2(1525)$, there is very little robust quantitative
information available that can directly be compared to theory. The
most restrictive datum appears to be the ratio of widths of $f_J(2220)$ to
$\pi\pi$ and $K\bar{K}$, which we shall analyse below. Of the qualitative
data available, the observation or non--observation of various 
isoscalar tensor mesons in $\phi\phi$ is of particular interest, 
as the $\phi\phi$ decay can arise for connected decay
only from the $s\bar{s}$ components of states. We shall not make 
exhaustive decay predictions, but restrict to the cases mentioned in a 
genuine attempt to confront our picture with experiment. However,
we first analyse the OZI forbidden decay $f^{'}_2(1525)\rightarrow\pi\pi$ 
which is zero in models where the state has only one valence component.

\subsection{$f^{'}_2(1525)\rightarrow\pi\pi$ \plabel{f1525}}

For the $13\times 13$ solutions 1a, 1b, 2a and 2b we find
$\Gamma(f^{'}_2(1525)\rightarrow\pi\pi) = 1.6 (1.4), 1.2 (1.1), 1.0$ $(0.9),
0.7 (0.7)$ MeV, with relativistic phase space\footnote{
In accordance with ref. \protect\cite{biceps} we use a slightly higher
pair creation constant for low--mass states. For 
$\Gamma(f_2(1275)\rightarrow\pi\pi)$,
with $f_2(1275)$ purely 1P $n\bar{n}$, this gives 160 MeV, in perfect
agreement the experimental $157\pm 4$ MeV \protect\cite{pdg98}.} listed first.
Solution 2b is in best agreement with the experimental value 
$0.60 \pm 0.12$ MeV. 
The two $9\times 9$ solutions gives the same results as their
$13\times 13$ counterparts. 
As seen in (\ref{val1}),(\ref{val2}), 
the valence content of the $f_2^{'}(1525)$ is 
such that its 1P and 2P $n\bar{n}$ components are in destructive 
interference (they have opposite signs), which will result in the suppression
of the $\pi \pi $ decay mode of this state.  
Furthermore, for components higher in mass than 2P, the valence content
has the same sign as the 2P component, leading to further suppression.
To be specific, we illustrate this for solution 2b. 
$\Gamma(f^{'}_2(1525)\rightarrow\pi\pi) = 14$ and $4$ MeV if only 1P, and
1P and 2P components are included. The width remains above 1.5 MeV as long
as not all of the 1P, 2P, 1F and 3P components are included.

We have thus provided the first quantitative understanding of the 
process $f^{'}_2(1525)\rightarrow\pi\pi$. This demonstrates
that the techniques of both the mass matrix and $^3P_0$ decay analysis
yield predictions consistent with experiment, motivating their continued
use. Remarkably, the decay $f^{'}_2(1525)\rightarrow\pi\pi$ can only be
understood when at least four different components of $f^{'}_2(1525)$ are
included. 
It is also apparent that there is no need to postulate a 
non--connected
decay mechanism, whereby primitive $s\bar{s}$ components would directly 
decay to $\pi\pi$. Because $f^{'}_2(1525)$ is dominantly $s\bar{s}$, such
processes must be small indeed. 
\question{exact valence formulae}

\subsection{$R\equiv\Gamma(f_J(2220)\rightarrow\pi^+\pi^-)/
\Gamma(f_J(2220)\rightarrow K^+K^-)\;\;\;  (J=2)$}

The main distinguishing characteristic of the $f_J(2220)$ is its 
remarkably narrow
total width of $23^{+8}_{-7}$ MeV \cite{pdg98} for a state that 
can decay via numerous
decay modes \cite{blundell}. Not even the existence of $f_2(2220)$ 
is well--established
\cite{review} as the narrow peak sit on a variety of empirical backgrounds,
for which there is no explanation \cite{klempt}, 
so that the peak might be a statistical fluctuation.
Moreover, broader states in the same mass region have been reported: 
JETSET sees an $f_2$
at $2231\pm 2$ MeV with a width of $70\pm 10$ MeV \cite{palano}. 
An $f_2$ at $2240\pm 40$ MeV
with a width of $170\pm 50$ MeV \cite{highten}, and at $2210\pm 45$ MeV 
with width
$260\pm 45$ \cite{buggf} have also been reported.
 It is possible to understand current data on $f_J(2220)$ if one does 
not take it to be
narrow \cite{klempt}. If the $f_J(2220)$ is not narrow, none of 
the ``indicators''
of its glueball nature can be sustained, for example its non--observation in
$\gamma\gamma$ collisions \cite{review} simply follows from its
wideness, and its coupling to gluons in $J/\Psi$ radiative decay becomes
compatible to conventional mesons if $J=2$ \cite{farrar}. 

As seen in (\ref{val1}),(\ref{val2}), the largest non--glue components of the 
$f_J(2220),$ 3P and 1F $s\bar{s},$ are in destructive 
interference, and remain so in the $13\times 13$ solutions. 
This also tends to be true for the largest $n\bar{n}$ components in
the $13\times 13$ solutions. One may think
this will result in the suppression of the decay
modes of this state, making its total width consistent with the
tiny experimental value. Evaluating the total width
of $f_J(2220)$, with $J=2$, to $\pi\pi$ and $K\bar{K}$, for all the
$13\times 13$ solutions, and phase space conventions, we obtain
$20-150$ MeV. 
It is evident that the tiny total width can not be sustained in our model, and
it likely is to be a challenge to any model in which the physical glueball 
has non--negligible mixing with mesons \cite{blundell}.
The individual partial widths to $\pi\pi$ and $K\bar{K}$ are also inconsistent
with experimental bounds that assume a narrow $f_J(2220)$ \cite{review}.

It is often claimed that $f_J(2220)$ has a flavour democratic decay pattern
expected for a pure glueball \cite{bai96}, 
whereby $R=1$ without phase space included, and
$R=1.7$ with phase space included. However, na\"{\i}ve flavour factors
give that a pure $n\bar{n}$ and $s\bar{s}$ state should have, 
without phase space included, $R=4,0$ respectively. Thus a mixture between
$n\bar{n}$ and $s\bar{s}$ can also look flavour democratic.
For the $13\times 13$ solutions 1a, 1b, 2a and 2b we find
$R= 0.6, 0.7, 0.4, 0.5$ respectively, independent of phase space
conventions. The two $9\times 9$ solutions gives the same results as their
$13\times 13$ counterparts. Although these values of $R$ do not
represent flavour democratic decay, they are all consistent with 
experiment \cite{bai96}, which possesses large error bars.
This is true for solutions 1a and 1b where $f_J(2220)$ is the physical
glueball, and for the other solutions.

When decays are to final S--wave mesons, i.e. 
$\pi,\eta,K,\rho,\omega,K^{\ast},\eta^{'}$ or $\phi$, one almost always 
finds that the decay amplitudes decrease sharply as the decaying component
is progressively radially excited \question{Just for S--wave?}.
The same is true as the decaying component is orbitally excited.
This has the consequence that although a physical state may have
a dominantly excited component, its decay dominantly proceeds 
through a lower excited component.
This means that na\"{\i}ve quark model calculations that assign a single
component to an excited 
state \cite{biceps,kokoski87} might be completely unreliable.
One would {\it a priori} expect this situation to be worst in $J^{PC}$ 
sectors where low--lying glueballs are present, i.e. $J^{PC} = 0^{++},
2^{++}$ and $0^{-+}$ \cite{morn}. 
We illustrate the phenomenon by analysing $f_J(2220)\rightarrow\pi\pi$ and
$K\bar{K}$ for the $13\times 13$ solutions. Although $f_J(2220)$ is
never dominantly 1P, this contribution is always one of the dominant
ones. One tends to finds that half of the width can be found by including
only the 1P and 2P contributions, even though the state may be
dominantly 1F and 3P\question{more simulations}.

\subsection{Decays to $\phi\phi$}

We shall analyse the decay
of various resonances to $\phi\phi$, in an attempt to understand the data,
which claim that $f_2(2010),f_2(2300)$ and $f_2(2340)$ have been seen in 
$\phi\phi$ in $\pi p$ collisions \cite{pdg98}. 
There is also preliminary evidence for an $f_2$ at $\sim 2231\pm 2$ MeV 
in $\phi\phi$ \cite{palano}\question{CHECK 3D WAVE}.
We bear in mind that the production process can alter conclusions
made based on studying decays.

The results are in Table \ref{tab}. From phase space considerations, it
is especially surprising  that it is
possible for $f_2(2150)$ to have a smaller width to $\phi\phi$ than
$f_2(2010)$. This is even more surprising, given that $f_2(2010)$ is
dominantly $n\bar{n}$ and $f_2(2150)$ is dominantly 1F $s\bar{s}$. The
resolution of the paradox is that 1F and 2F does not decay to
$\phi\phi$ in the dominant S--wave in the $^3P_0$ model \cite{biceps}. 
We note that solutions 2a and 2b are consistent 
with states unambiguously observed in $\phi\phi$.

\begin{table}[t]
\begin{center}
\begin{tabular}{|c||c|c|c|c|c|}
\hline 
$13\times 13$ solution & $f_2(2010)$& $f_2(2150)$& $f_J(2220)$
& $f_2(2300)$& $f_2(2340)$  \\
\hline \hline 
1a & 3(3) &  7(5) &  17(12) &  23(16) &  7(4) \\  
1b & 3(2) &  5(4) &  12(9) &  17(12) &  8(5) \\  
2a & 18(15) &  3(3) &  0.6(0.5) &  4(3) &  9(6) \\  
2b & 12(11) &  1.7(1.4) &  0.7(0.5) &  7(5) &  2(2) \\  
\hline 
\end{tabular}
\caption{\plabel{tab}
Decay widths to $\phi\phi$ in MeV for various solutions. 
Relativistic phase space
is used first, with mock meson phase space in brackets. 
The two $9\times 9$ solutions give almost identical results to their
$13\times 13$ counterparts. $f_2(2010)$ is taken
to have a mass at the upper end of the experimental 
range \protect\cite{pdg98}. } 
\end{center}
\end{table}

\question{Most states above 2 GeV have either $\phi\phi$ or $\pi\pi$ 
coupling in PDG.
New $p\bar{p}$ results. Which solution this prefers?}

\question{PDG assigns nnbar  2P to f2(1810) and ssbar 2P to f2(2010)}

\question{Evidence for low P-wave radial mass may indicate early breaking of 
string tensor (see G.Bali ``Overview of lattice QCD''}

\section{Salient features}
\question{what can we predict to test our approach?}

We showed that Schwinger--type mass formulae can be obtained when we
restrict to glueball-- (hybrid) meson mixing. With some physical isovector
and isoscalar masses known, these formulae can predict unknown masses
and couplings. The utility of this new analysis technique was demonstrated in 
the tensor sector.

It has been shown that in order to understand the decay $f_2'(1525)\rightarrow
\pi \pi ,$ one has to consider more than the 1P $n\bar{n}$ component. This 
implies that the use of a $2\times 2$ mixing formula where the 
physical states are linear combinations of $n\bar{n}$ and $s\bar{s}$ 
components can be inadequate. 

In our approach the physical glueball  is {\it a priori} narrower than mesons 
due to the large glueball component, which is
taken not to decay. However, as shown for the $f_J(2220)$, the 
physical glueball is not unusually narrow, because of the presence of 
significant meson components, contrary to the perturbative QCD claim that
glueball mixing with 1F mesons is tiny \cite{pertqcd1}. Also, there is no 
reason to expect a flavour democratic decay pattern for the physical 
glueball. Experimentally, this is already clear for the scalar glueball 
\cite{review,scalar}. It is a common myth that primitive glueballs are 
narrow and decay flavour democratically. These notions arise from perturbative 
QCD, which, paradoxically, has been argued to be
valid for the tensor glueball but not for the scalar glueball
\cite{pertqcd2}. However, glueballs beyond 3 GeV can be narrow \cite{bur98}.
A more reliable glueball signature may be glue--rich production.

\vspace{1cm}

{\bf Acknowledgements}

We had useful discussions with D.V. Bugg, F.E. Close, A. Kirk, 
W. Lee and B.-S. Zou.

\appendix
\catcode`\@=11 \@addtoreset{equation}{section}

\renewcommand{\theequation}{A.\arabic{equation}}  
\section{Appendix: $13\times 13$ mass matrix \plabel{appc}}

The  $13\times 13$ mass matrix for solution 2b is 

{\footnotesize
\beqn
\left(
\begin{array}{ccccccccccccc}
2.25 & 0.08 & 0.08\;\sqrt{2} & 0.08 & 0.08\;\!\sqrt{2} & 0.13 & 0.13\;\!
\sqrt{2} & 0.14 & 0.14\;\!\sqrt{2} & 0.08 & 0.08\;\!\sqrt{2} & 0.12 & 0.12\;
\!\sqrt{2}  \\
0.08 & 2.55 & 0 & 0 & 0 & 0 & 0 & 0 & 0 & 0 & 0 & 0 & 0  \\
0.08\;\!\sqrt{2} & 0 & 2.35 & 0 & 0 & 0 & 0 & 0 & 0 & 0 & 0 & 0 & 0  \\
0.08 & 0 & 0 & 2.5 & 0 & 0 & 0 & 0 & 0 & 0 & 0 & 0 & 0  \\
0.08\;\!\sqrt{2} & 0 & 0 & 0 & 2.3 & 0 & 0 & 0 & 0 & 0 & 0 & 0 & 0  \\
0.13 & 0 & 0 & 0 & 0 & 2.27 & 0 & 0 & 0 & 0 & 0 & 0 & 0  \\
0.13\;\!\sqrt{2} & 0 & 0 & 0 & 0 & 0 & 2.05 & 0 & 0 & 0 & 0 & 0 & 0  \\
0.14 & 0 & 0 & 0 & 0 & 0 & 0 & 2.15 & 0 & 0 & 0 & 0 & 0  \\
0.14\;\!\sqrt{2} & 0 & 0 & 0 & 0 & 0 & 0 & 0 & 1.94 & 0 & 0 & 0 & 0  \\
0.08 & 0 & 0 & 0 & 0 & 0 & 0 & 0 & 0 & 1.95 & 0 & 0 & 0  \\
0.08\;\!\sqrt{2} & 0 & 0 & 0 & 0 & 0 & 0 & 0 & 0 & 0 & 1.66 & 0 & 0  \\
0.12 & 0 & 0 & 0 & 0 & 0 & 0 & 0 & 0 & 0 & 0 & 1.55 & 0  \\
0.12\;\!\sqrt{2} & 0 & 0 & 0 & 0 & 0 & 0 & 0 & 0 & 0 & 0 & 0 & 1.318
\end{array}
\right) .
\eeqn
}
The physical masses are 
\beqn
2.67,\;2.53,\;2.47,\;2.34,\;2.29,\;2.23,\;2.12,
\;2.01,\;1.95,\;1.81,\;1.64,\;1.52,\;1.28,
\eeqn
and the valence content of the physical states is
{\footnotesize
\beqn
\left(
\begin{array}{ccccccccccccc}
0.65 & 0.45 & 0.23 & 0.32 & 0.20 & 0.21 & 0.19 & 0.18 & 0.18 & 
0.07 & 0.07 & 0.07 & 0.08  \\
0.19 & -0.84 & 0.12 & 0.46 & 0.09 & 0.09 & 0.07 & 0.07 & 0.06 & 
0.02 & 0.02 & 0.02 & 0.03  \\
0.27 & -0.29 & 0.25 & -0.82 & 0.18 & 0.18 & 0.12 & 0.12 & 0.10 & 
0.04 & 0.04 & 0.04 & 0.04  \\
0.11 & -0.04 & -0.88 & -0.05 & 0.37 & 0.23 & 0.07 & 0.09 & 0.06 & 
0.02 & 0.02 & 0.02 & 0.02  \\
0.08 & -0.02 & -0.15 & -0.03 & -0.79 & 0.57 & 0.06 & 0.08 & 0.05 & 
0.02 & 0.01 & 0.01 & 0.01  \\
0.23 & -0.06 & -0.21 & -0.07 & -0.36 & -0.69 & 0.24 & 0.43 & 0.16 & 
0.07 & 0.05 & 0.04 & 0.04  \\
-0.17 & 0.03 & 0.08 & 0.04 & 0.11 & 0.15 & -0.44 & 0.83 & -0.19 & 
-0.08 & -0.04 & -0.04 & -0.04  \\
-0.18 & 0.03 & 0.06 & 0.03 & 0.07 & 0.09 & 0.74 & 0.17 & -0.54 & 
-0.26 & -0.06 & -0.05 & -0.04  \\
-0.02 & 0.00 & 0.00 & 0.00 & 0.00 & 0.01 & 0.03 & 0.01 & -0.38 & 
0.92 & -0.01 & -0.00 & -0.00  \\
-0.43 & 0.05 & 0.09 & 0.05 & 0.10 & 0.12 & 0.32 & 0.17 & 0.64 & 
0.24 & -0.33 & -0.20 & -0.15  \\ 
-0.20 & 0.02 & 0.03 & 0.02 & 0.03 & 0.04 & 0.09 & 0.05 & 0.13 &
0.05 & 0.92 & -0.28 & -0.10  \\
-0.22 & 0.02 & 0.03 & 0.02 & 0.03 & 0.04 & 0.08 & 0.05 & 0.10 & 
0.04 & 0.18 & 0.93 & -0.18  \\
-0.23 & 0.01 & 0.02 & 0.01 & 0.03 & 0.03 & 0.05 & 0.04 & 0.07 & 
0.03 & 0.07 & 0.10 & 0.96
\end{array}
\right) .
\eeqn
}

Solution 2b is similar to the $9\times 9$ solution 2 in Eqs. 
(\ref{mass2})-(\ref{val2}), but is not the $13\times 13$ 
counterpart of it, as evidenced
by the different couplings of the common primitive states and 
the different primitive glueball mass. 

We now study the stability of the solution under parameter changes. 
When we change the primitive
1F $n\bar{n}$ and $s\bar{s}$ masses upwards by 50 and 40 MeV 
respectively, reflecting
our lack of knowledge of these parameters, the physical masses all 
remain consistent
with experiment, except for $f_J(2220)$, which is 6 MeV higher than 
the experimental
mean \cite{pdg98}. Valence contents of states far away in mass from 1F remain 
essentially constant. The largest changes are found for $f_2(2010)$, where 
valence contents less than $0.2$ change on average by $70\%$. The 
dominant content 
changes by $1\%$ and the next most dominant by $20\%$. A further 
change of the 2P $s\bar{s}$
mass downwards by 30 MeV yields the largest  changes for the valence content of
$f_2(1950)$, by similar amounts as before.   





\renewcommand{\theequation}{B.\arabic{equation}}  
\section{Appendix: Motivation for mass matrix \plabel{appa}}

If an appropriate basis is chosen, the hamiltonian for $n$ isovector states
up to a certain mass
can  be taken to  be a diagonal $n\times n$ matrix
${\cal N}\equiv \mbox{diag}(N_1,N_2,\ldots ,N_n)$, where the entries are 
real and
positive. These entries are identified with the masses of the physical
states in the experimental isovector spectrum. Note that no assumption
is made about the nature of the states, e.g. conventional or hybrid meson or
four--quark state. In a world of only $u,d$ quarks, this work assumes
degeneracy of isovector and isoscalar states, motivated in ref. 
\cite{sch}. The hamiltonian for the isoscalar states
is simply the $(n+1)\times (n+1)$ matrix 

\beqn\plabel{m1} \left(
\barr{cc} G & \sqrt{2} g \\ \sqrt{2} g^T & {\cal N} \earr
\right) \eeqn
where a new state, which cannot exist in the isovector sector, the 
glueball, has been added. $G$ is the primitive glueball mass, and
$g$ is a n--dimensional row of (real) couplings of either 
$u\bar{u}$ or $d\bar{d}$. These couplings are
${\cal O}(\frac{1}{\sqrt{N_c}})$ in the large number of colours $N_c$ 
expansion of QCD. Of course, any number of extra glueballs can in principle 
be added \cite{sch}. \question{futher motivation from Schwinger paper}

If the strange quark is also incorporated, the isoscalar matrix becomes
(\ref{mass}). 
Here the matrix element between $n\bar{n}$ states $i$ and $j$ ($i\neq j$) 
is zero
because of the diagonality of ${\cal N}$ in (\ref{m1}). The strange quark
states are taken to be heavier analogues of the light ones, so that the
mixing between $s\bar{s}$ states $i$ and $j$ ($i\neq j$) is zero, and 
the diagonal entries
contain the masses  $S_i$ of the strange quark states.  The remaining
possible mixing is between $n\bar{n}$ state $i$ and $s\bar{s}$ state $j$, 
which is ${\cal O}(\frac{1}{{N_c}})$ \cite{scalar}
and was found tiny in recent lattice QCD simulations \cite{LW1}, 
and is neglected in this work.
The strange quark states are assumed to couple in the same way to the
glueball as $u\bar{u}$ and $d\bar{d}$, which is the SU(3) limit. 
For ground state
isoscalar scalar mixing, lattice QCD obtains the ratio of $u\bar{u}$ 
and $d\bar{d}$ to $s\bar{s}$ glueball coupling to be
$1.198 \pm 0.072$ \cite{LW1}, while the SU(3) limit is 1.

\renewcommand{\theequation}{C.\arabic{equation}}  
\section{Appendix: Decay formalism \plabel{appd} }

The mass matrix in Eq. \ref{mass} can be viewed as forming part of 
a hamiltonian $H$ that describes an effective theory.
This part of the hamiltonian is diagonalized to yield the
physical states. One can then  {\it a posteriori} 
 add to the hamiltonian a part that describes
coupling to a decay channel. 

\beqn
H = g^{\ast}Gg + \sum_{i=1}^n (s_i^{\ast} S_i s_i + n_i^{\ast} N_i n_i) +
(\sum_{i=1}^n g_i g^{\ast} (s_i + \sqrt{2} n_i) + c.c.) + 
(\gamma^g g^{\ast} + \sum_{i=1}^n 
(\gamma^s_i s_i^{\ast} + \gamma^n_i n_i^{\ast} )) (BC)
\eeqn
where $G, N_i$ and $S_i$ are the masses and $g_i$ the couplings in Eq. 
\ref{mass};
 $g, n_i, s_i$ the corresponding 
primitive glueball, $n\bar{n}$ and $s\bar{s}$ meson fields; and 
$\gamma^g,\gamma^n_i,\gamma^s_i$ represent the couplings of the 
respective primitive states to
the decay channel field $(BC)$. Spin indices for the primitive tensor states
have been suppressed in $H$, e.g. $g^{\ast}Gg$ stands for
$g^{\ast}_{\mu\nu}Gg^{\mu\nu}$, where $g^{\mu\nu}$ is a symmetric and
traceless Lorentz tensor. 

Now write the hamiltonian in shortened notation

\beqn
H = {\bf m}^{\dagger} {\cal M} {\bf m} + {\bf m}^{\dagger} {\bf \gamma}(BC)
\hspace{1cm} {\bf m}\equiv (g, s_1, n_1 \ldots s_n, n_n)^T 
\hspace{1cm} {\bf \gamma}\equiv (\gamma^g, \gamma^s_1, \gamma^n_1 
\ldots\gamma^s_n, \gamma^n_n)^T 
\eeqn
where $\cal M$ is defined as the matrix (\ref{mass}). Since $\cal M$ 
is real and symmetric
it can be diagonalized by use of the (real) 
orthogonal ($\Omega^{-1}=\Omega^{T}$) valence content matrix 
$\Omega$, to yield
the diagonal mass matrix of (real) eigenvalues (physical masses) 
${\cal M}_D = \Omega{\cal M}\Omega^{-1}$. 
The eigenvectors (physical states) are $\tilde{\bf m} = \Omega {\bf m}$,
where ${\bf m}$ denotes the primitive states. 
The $j^{th}$ row in $\Omega$ gives the valence content of the 
$j^{th}$ physical state. Several practical examples of the 
diagonalization procedure can be found in 
this work. For example, for the ${\cal M}$ in (\ref{mass1}), 
${\cal M}_D$ is the matrix with diagonal entries (\ref{phys1}), and
$\Omega$ is (\ref{val1}).

The hamiltonian becomes

\beqn
H=\tilde{\bf m}^{\dagger} {\cal M}_D \tilde{\bf m} + \tilde{\bf m}^{\dagger} 
\Omega {\bf \gamma}(BC)
\eeqn
The first term was discussed in detail in ref. \cite{ohio}.
The second term shows clearly that in order to calculate the decay amplitude
of a physical state to (BC), it is necessary to add the decay amplitudes
of all its primitive components, weighted by their valence content.
This was used, but not explicitly demonstrated in ref. \cite{ohio}.

\end{document}